\documentclass[aps,prd,amsmath,amssymb,preprintnumbers,nofootinbib,a4paper,11pt]{revtex4-1}
\pdfoutput=1
\usepackage{amsthm}

\usepackage{graphicx}
	\graphicspath{{./NewFig/}}
\usepackage{subcaption}

\usepackage{url}
\usepackage[bookmarks, pagebackref=false]{hyperref}
\usepackage{color}
	\definecolor{rossoCP3}{cmyk}{0,.88,.77,.40}
		\definecolor{graa}{rgb}{0.8,0.8,0.8}
		\definecolor{blaa}{rgb}{0.2,0.2,0.6}
		\hypersetup{
			colorlinks,
			bookmarksopen,
			bookmarksnumbered,
			citecolor=blaa, 		
			linkcolor=rossoCP3,	
			urlcolor=rossoCP3,			
			}
\usepackage{dcolumn}
\usepackage{bm}
\usepackage{bbm}
\usepackage{pxfonts}

\usepackage{epsfig}
\usepackage{placeins}

\usepackage{cleveref}

\usepackage[ margin=5pt, font=small,labelfont=bf, justification=raggedright]{caption}

\usepackage{youngtab}
\usepackage{slashed}

\newcommand{\beq}{\begin{eqnarray}}
\newcommand{\eeq}{\end{eqnarray}}

\newcommand{\bmp}{\noindent\begin{minipage}{16cm}}
\newcommand{\emp}{\end{minipage}\vskip 7mm} 

\def\lsim{\mathrel{\rlap{\lower4pt\hbox{\hskip1pt$\sim$}}
    \raise1pt\hbox{$<$}}}                
\def\gsim{\mathrel{\rlap{\lower4pt\hbox{\hskip1pt$\sim$}}
    \raise1pt\hbox{$>$}}}                

\begin{document}
\title{\Large   Properties of the $\epsilon$-Expansion, Lagrange Inversion and Associahedra and the $O(1)$ Model. }
\author{Thomas A. Ryttov$^{\varheartsuit}$}\email{ryttov@cp3.sdu.dk}
\affiliation{
$^{\varheartsuit}${  \rm CP}$^{\bf 3}${ \rm-Origins}, University of Southern Denmark, Campusvej 55, 5230 Odense M, Denmark}

\begin{abstract}
We discuss properties of the $\epsilon$-expansion in $d=4-\epsilon$ dimensions. Using Lagrange inversion we write down an exact expression for the value of the Wilson-Fisher fixed point coupling order by order in $\epsilon$ in terms of the beta function coefficients. The $\epsilon$-expansion is combinatoric in the sense that the Wilson-Fisher fixed point coupling at each order depends on the beta function coefficients via Bell polynomials. Using certain properties of Lagrange inversion we then argue that the $\epsilon$-expansion of the Wilson-Fisher fixed point coupling equally well can be viewed as a geometric expansion which is controlled by the facial structure of associahedra. We then write down an exact expression for the value of anomalous dimensions at the Wilson-Fisher fixed point order by order in $\epsilon$ in terms of the coefficients of the beta function and anomalous dimensions. 

We finally use our general results to compute the values for the Wilson-fisher fixed point coupling and critical exponents for the scalar $O(1)$ symmetric model to $O(\epsilon^7)$. 
\end{abstract}

\maketitle

\section{Introduction}

The scalar $O(N)$ symmetric theory $(\phi^i\phi^i)^2$ in $d=4-\epsilon$ dimensions is extremely rich in the number of physical phenomena that it can accommodate and describe. For instance for $N=0$ it coincides with the critical behavior of polymers (self-avoiding walks) while for $N=1$ it lies in the same universality class as liquid-vapor transitions and uniaxial magnets (Ising). For $N=2$ the theory describes the superfluid transition of liquid helium close to the $\lambda$-point ($XY$). For $N=3$ it describes the critical behavior of isotropic ferromagnets (Heisenberg). There are many more examples of critical phenomena that are described by the scalar $O(N)$ symmetric model and its close relatives for which the reader can find a more complete discussion in \cite{Pelissetto:2000ek}. It suffices to say that further understanding of the theory is of great importance.

For these reasons much effort has been put into computing various renormalization group functions in $d=4-\epsilon$ dimensions to higher and higher order in the coupling. The three and four loop calculation \cite{Brezin:1976vw,Kazakov:1979ik} as well as the five loop calculation \cite{Gorishnii:1983gp,Chetyrkin:1981jq,Kazakov:1984km,LeGuillou:1985pg,Kleinert:1991rg,Guida:1998bx} long stood as the state-of-the-art of higher loop calculations in $d=4-\epsilon$ dimensions. But recently the six loop calculations initiated in \cite{Batkovich:2016jus,Kompaniets:2016hct} were brought to the end in \cite{Kompaniets:2017yct} for general $N$. In addition an impressive seven loop calculation in the special case of $N=1$ has appeared \cite{Schnetz:2016fhy}. These now constitute the highest loop order calculations of the renormalization group functions which include the beta function, the field anomalous dimension and the mass anomalous dimension. 

At sufficiently small  $\epsilon$ the one loop beta function possesses a non-trivial zero \cite{Wilson:1971dc}. This is the Wilson-Fisher fixed point. At this fixed point anomalous dimensions and their associated critical exponents are expressed as power series in $\epsilon$. At a fixed point these critical exponents are scheme independent physical quantities. For general $N$ and using the six loop calculations they can be found in analytical form to $O(\epsilon^5)$ in the attached Mathematica file in \cite{Kompaniets:2017yct}. When combining these results with resummation techniques they agree well with experiments \cite{LeGuillou:1977rjt,LeGuillou:1979ixc,LeGuillou:1985pg,Kompaniets:2017yct}. We also note several recent complimentary studies of scalar field theories in various dimensions \cite{Rychkov:2018vya,Hogervorst:2015akt,Rychkov:2015naa,Fei:2014yja,Fei:2014xta,Rattazzi:2008pe,ElShowk:2012ht,El-Showk:2014dwa,Kos:2015mba,Kos:2016ysd,Hellerman:2015nra,Arias-Tamargo:2019xld,Watanabe:2019pdh,Badel:2019oxl,Sen:2015doa,Gliozzi:2017hni,Liendo:2017wsn,Dey:2017oim,Banerjee:2019jpw,Litim:2016hlb,Juttner:2017cpr}.

In this work we do not attempt to extend any of the higher loop calculations. We will instead entertain ourselves by providing explicit and closed form expressions for the fixed point value of the coupling and any anomalous dimension or critical exponent at the Wilson-Fisher fixed point to all orders in $\epsilon$. The closed form all orders expressions are then functions of all the coefficients of the beta function and anomalous dimension (which still need to be calculated). The main feature of the Wilson-Fisher fixed point is the fact that it is an expansion in the first coefficient (which is $\epsilon$) of the beta function in $d=4-\epsilon$. As we will see this allows us to use the Lagrange inversion theorem to formally find the zero of the beta function to all orders in $\epsilon$.

Given the formal power series $f(x) = \sum_{i=1}^{\infty} c_i x^i$ one might ask whether it is possible to find the coefficients $d_i$ in the power series  $g(y) = \sum_{i=1}^{\infty} d_i y^i$ where $f$ and $g$ are each others compositional inverses $g(f(x)) = x$ and $f(g(y) ) = y$. The Lagrange inversion theorem provides a procedure for doing to this and for the first few orders one finds $d_1 =1/c_1 $, $d_2=-c_2/c_1^3$, $d_3 = (2c_2^2 - c_1c_3)/c_1^5$, etc. There is also a general closed form expression for $d_i$ at any order which is given in terms of Bell polynomials. This can be found for instance in \cite{MorseFeshbach}. 

In our case where we study critical phenomena via the $\epsilon$-expansion it will be more convenient for us to phrase the question of power series inversion in a slightly different but equivalent way. Instead of asking for the compositional inverse of $f(x) = \sum_{i=1}^{\infty} c_i x^i$ we will ask for the zero of $ \tilde{f}(x) $ where $\tilde{f}(x) \equiv -a+f(x)= - a+  \sum_{i=1}^{\infty} c_i x^i$. The zero of $\tilde{f}(x)$ we are looking for should then be given as a power series in the first coefficient $a$ of $\tilde{f}(x)$ and be written as $x = \sum_{i=1}^{\infty}d_i a^i$. Of course one again finds $d_1 = 1/c_1$, $d_2 = - c_2/c_1^3$, $d_3 = (2c_2^2 - c_1c_3)/c_1^5$, etc as above. It is more natural for us to consider Lagrange inversion in this way. Further details will follow below where we will see the coefficients $d_i$ presented in both a combinatoric and geometric language within the physical setting of critical phenomena and the $\epsilon$-expansion.

The use of Lagrange inversion of power series in physics has a beautiful history in celestial mechanics and originates in the study of the restricted Newtonian three-body problem. Here one has to solve a quintic equation $0 = a+\sum_{i=1}^{5} c_i x^i $ in $x$ for three sets of coefficients in order to find the location of the three collinear Lagrange points $L_1$, $L_2$ and $L_3$. Although it is impossible to find general solutions in terms of radicals of a quintic equation it is possible to find a single real solution which is an expansion in the coefficient $a$ as $x = \sum_{i=1}^{\infty} d_i a^i $. One can in principle compute the coefficients $d_i$ to any desired order using the Lagrange inversion theorem. The success of the Lagrange method to find the locations of $L_1$, $L_2$ and $L_3$ is perhaps best illustrated by the fact that in the Earth-Sun system the Solar and Heliospheric Observatory (SOHO) occupies $L_1$ while the Wilkinson Microwave Anisotropy Probe (WMAP) occupies $L_2$.

The paper is organized as follows. In Section \ref{WF} we derive the value of the Wilson-Fisher fixed point coupling in terms of the beta function coefficients to all orders in $\epsilon$ and show that it  is given in terms of Bell polynomials. In Section \ref{associahedron} we press on and discuss how the $\epsilon$-expansion of the Wilson-Fisher fixed point coupling also can be understood as a geometric expansion controlled by associaheda. In Section \ref{anodims} we discuss anomalous dimensions while in Section \ref{O(1)} we compute the fixed point coupling and critical exponents to $O(\epsilon^7)$ for the $O(1)$ model. We conclude in Section \ref{conclusions}.

 \section{The Wilson-Fisher Fixed Point}\label{WF}

We write the beta function of the coupling $g$ as $\beta(g,\epsilon)$ and in $d=4 - \epsilon$ dimensions it is generally given as a formal power series expansion
\begin{eqnarray}
\beta (g,\epsilon) &=& -\epsilon g + \sum_{i=1}^{\infty} b_{i-1} g^{i+1} = -\epsilon g + b_0 g^2 + b_1 g^3 +\ldots
\end{eqnarray}
where $b_i$, $i=0,1,\ldots$ are the standard beta function coefficients. We are interested in the fixed points $\beta(g_*,\epsilon) = 0$ of the theory in $d= 4 - \epsilon$ dimensions. Clearly the trivial fixed point $g_*= 0$ always exists. At one loop the beta function also possesses a non-trivial fixed point $g_* = \frac{\epsilon}{b_0} $. This is the Wilson-Fisher fixed point.

We will be interested in the physics of the Wilson-Fisher fixed point and how to derive a closed form exact expression for the fixed point value of the coupling $g_*$ to all orders in $\epsilon$. First write the fixed point equation $\beta(g_*,\epsilon) = 0$ we want to solve as
\begin{eqnarray}\label{FPequation}\label{zero}
0 &=&- \tilde{\epsilon} + \sum_{i=1}^{\infty} \tilde{b}_{i-1} g_*^{i} = - \tilde{\epsilon} +   g_* + \tilde{b}_1 g_*^2 + \tilde{b}_2 g_*^3+ \ldots 
\end{eqnarray}
where for convenience we have chosen to normalize everything as $\tilde{\epsilon} = \frac{\epsilon}{b_0}$ and $\tilde{b}_{i} = \frac{b_i}{b_0}$ so that the coefficient of $g_*$ is unity. This is purely a matter of convenience. We want to find the zero $g_*(\tilde{\epsilon})$ which is a power series in $\tilde{\epsilon}$. As already explained this is encoded in the Lagrange inversion theorem. Finding $g_*(\tilde{\epsilon})$ amounts to determining the coefficients $g_i$ in the power series
\begin{eqnarray}
g_*(\tilde{\epsilon}) &=& \sum_{i=1}^{\infty} g_{i-1} \tilde{\epsilon}^i = g_0 \tilde{\epsilon} +g_1 \tilde{\epsilon}^2  + g_2 \tilde{\epsilon}^3+ \ldots
\end{eqnarray}
One way to do this is to first plug the ansatz for $g_*(\tilde{\epsilon})$ back into Eq. \ref{zero} and then expand again in $\tilde{\epsilon}$ to arrive at\footnote{An alternative method can again be found in \cite{MorseFeshbach}}
\begin{eqnarray}\label{expansion}
0  &=& - \tilde{\epsilon} + \sum_{j=1}^{\infty} \tilde{b}_{j-1} \left( \sum_{l=1}^{\infty} g_{l-1} \tilde{\epsilon}^l \right)^j    = \left( g_0-1\right)  \tilde{\epsilon} + \left( g_1 + g_0^2 \tilde{b}_1 \right) \tilde{\epsilon}^2 + \left(g_2 + 2 g_0g_1 \tilde{b}_1 + g_0^3\tilde{b}_2 \right) \tilde{\epsilon}^3 + \ldots + c_i \tilde{\epsilon}^i+ \dots \nonumber \\
\end{eqnarray}
In order to write an explicit closed form expression for the $i$'th coefficient $c_i$ we need to know what terms in the composite set of sums will contribute at $O(\tilde{\epsilon}^i)$. If we first look at the infinite sum $ \sum_{l=1}^{\infty} g_{l-1} \tilde{\epsilon}^l $ then clearly only a finite number of terms $g_0 \tilde{\epsilon} +g_1 \tilde{\epsilon}^2 +\ldots + g_{k-1} \tilde{\epsilon}^k$ for some $k \leq i$, can eventually contribute at $O(\tilde{\epsilon}^i)$. How large can $k$ be until the term $g_{k-1} \tilde{\epsilon}^k$ will no longer contribute at $O(\tilde{\epsilon}^i)$? Once we take the finite number of terms to the power $j$, the term that involves a single factor of $g_{k-1} \tilde{\epsilon}^k$ and is lowest order in $\tilde{\epsilon}$ is $(g_0 \tilde{\epsilon})^{j-1} g_{k-1} \tilde{\epsilon}^k $. So if this is to be of order $O(\tilde{\epsilon}^i)$ then clearly $k=i-j+1$. Using the multinomial formula we can therefore write
\begin{eqnarray}
 \left(g_0 \tilde{\epsilon} +g_1 \tilde{\epsilon}^2 +\ldots + g_{i-j} \tilde{\epsilon}^{i-j+1} \right)^j =   \sum_{j_1+\ldots + j_{i-j+1} = j} \frac{j!}{j_1 !\cdots j_{i-j+1}!} \left( g_0^{j_1} \cdots g_{i-j}^{j_{i-j+1}} \right)  \tilde{\epsilon}^{1j_1 + \ldots + (i-j+1) j_{i-j+1}}
\end{eqnarray}
where the sum is over all sequences $j_1,\ldots,j_{i-j+1}$. Precisely for $1j_1 + \ldots + (i-j+1) j_{i-j+1} = i$ we pick up the term of $O(\tilde{\epsilon}^i)$ and we can therefore read off the desired coefficient as
\begin{eqnarray}\label{coefficient}
c_i = \frac{1}{i!} \sum_{j=1}^i j! \tilde{b}_{j-1} B_{i,j} \left(1! g_0,\ldots, (i-j+1)!g_{i-j} \right)
\end{eqnarray}
We have here chosen to write the coefficient $c_i$ in terms of the Bell polynomials
\begin{eqnarray}
B_{i,j}(x_1,\ldots,x_{i-j+1}) = \mathop{\sum_{j_1 + \ldots + j_{i-j+1} =j}}_{1j_1 +\ldots  + (i-j+1) j_{i-j+1} =i}   \frac{i!}{j_1!\cdots j_{i-j+1}!}  \left(\frac{x_1}{1!}  \right)^{j_1} \cdots \left( \frac{x_{i-j+1}}{(i-j+1)!} \right)^{j_{i-j+1}} 
\end{eqnarray}
For completeness we note that it is also possible to find the coefficient by using the Fa{\`a} di Bruno formula for the $i$'th derivative of a composite function. However the above derivation is quite straightforward and perhaps not everyone is familiar with the Fa{\`a} di Bruno formula.\footnote{The Fa{\`a} di Bruno formula is \begin{eqnarray}
\frac{d^i }{d x^i} f(h(x)) &=& \sum_{j=1}^i f^{(j)} (h(x)) B_{i,j} \left( h'(x), h''(x), \dots, h^{(i-j+1)} (x)  \right)
\end{eqnarray}} Also for a discussion on Bell polynomials we refer the reader to \cite{Comtet}.

Bell polynomials are well known in combinatorics. They give a way to encode, in a polynomial, the partitioning of a set into non-empty, non-overlapping subsets. Assume we have a set of $i$ elements that we want to partition into $j$ non-empty, non-overlapping subsets. Each subset in the partition will contain some number of original elements $k$, which can be any $k=1,\ldots,i-j+1$. If there are $l$ such subsets and this specific partitioning can be done in $m$ ways then this is encoded as $mx_k^l$ in the Bell polynomial $B_{i,j}(x_1,\ldots,x_{i-j+1})$.\footnote{An example: Take a set with $i=5$ elements $\{a,b,c,d,e\}$ that we want to partition into $j=3$ subsets. First this can be done as $\{\{ a,b \},\{ c,d\}, \{e\}\}$ where $l=1$ subsets contain $k=1$ elements and $l=2$ subsets contain $k=2$ elements. This partitioning can be done in $m=15$ different ways. The partitioning into three subsets can also be done as $\{ \{a,b,c \},\{d\},\{e\} \}$ where $l=2$ subsets contain $k=1$ elements and $l=1$ subsets contain $k=3$ elements. This partitioning can be done in $m=10$ different ways. All this is precisely encoded in the Bell polynomial $B_{5,3}(x_1,x_2,x_3) =  15 x_1x_2^2+ 10x_1^2x_3$ }

The fact that the coefficients $c_i$ are combinatorial in the $\tilde{b}_i$'s and $g_i$'s should of course not come as a surprise since they are the $i$'th derivative of a composite function. Now it is important to realize that the expansion in Eq. \ref{expansion} should hold for varying $\tilde{\epsilon}$ so we can equate coefficients order by order in $\tilde{\epsilon}$. The coefficient of $\tilde{\epsilon}$ allows to solve for $g_0$, the coefficient of $\tilde{\epsilon}^2$ allows to solve for $g_1$ (already knowing $g_0$) and the coefficient of $\tilde{\epsilon}^3$ allows to solve for $g_2$ (already knowing $g_0$ and $g_1$). This can be done order by order to any order. In general the coefficient $c_i$ of $\tilde{\epsilon}^i$ depends on $g_0\ldots,g_{i-1}$ and is linear in $g_{i-1}$ so one can always solve for it (already knowing $g_1,\ldots,g_{i-2}$). It is linear in $g_{i-1}$ since the Bell polynomials satisfy the following identity $B_{i,1}(1!g_0,\ldots,i!g_{i-1}) = i! g_{i-1} $ which comes about in the $j=1$ term in the sum in $c_i$. Order by order we then find
\begin{eqnarray}
g_0 &=& 1\label{g0combinatoric} \\
g_1 &=& - \tilde{b}_1 \label{g1combinatoric} \\
g_2 &=& 2\tilde{b}_1^2 -\tilde{b}_2 \label{g2combinatoric} \\
g_3 &=& -5\tilde{b}_1^3 + 5 \tilde{b}_1\tilde{b}_2 -  \tilde{b}_3 \label{g3combinatoric} \\
g_4 &=& 14 \tilde{b}_1^4 - 21 \tilde{b}_1^2 \tilde{b}_2 + 3 \tilde{b}_2^2 + 6 \tilde{b}_1 \tilde{b}_3 - \tilde{b}_4  \label{g4combinatoric}\\
&\vdots& \nonumber \\
g_{i-1} &=& \frac{1}{i!} \sum_{j=1}^{i-1}  (-1)^j \tilde{i}  B_{i-1,j} \left(1! \tilde{b}_1 ,\ldots, (i-j)!\tilde{b}_{i-j}  \right)  \label{gcombinatoric}
\end{eqnarray}
where $\tilde{i} = i(i+1)\cdots (i+j-1)$. This gives us every coefficient $g_{i-1}$ and hence the value of the fixed point coupling to all orders in $\epsilon$ in terms of the beta function coefficients
\begin{eqnarray}\label{eq:g}
g_*(\epsilon) = \sum_{i=1}^{\infty} g_{i-1} \left( \frac{\epsilon}{b_0} \right)^i  \ , \qquad g_{i-1} = \frac{1}{i!} \sum_{j=1}^{i-1}  (-1)^j \tilde{i}  B_{i-1,j} \left(\frac{1! b_1}{b_0} ,\ldots, \frac{(i-j)! b_{i-j}}{b_0}  \right)   \ , \qquad g_0=1
\end{eqnarray}
This is a very concise and compact formula for the fixed point value to all orders in $\epsilon$ and is combinatoric in the sense that it is given in terms of the Bell polynomials. 

Note that for a given $i$ the coefficient $g_{i-1}$ depends on the $i$ loop beta function coefficients $b_0,\ldots,b_{i-1}$ only and does not receive corrections from higher orders. So for the scalar $O(N)$ symmetric model where the beta function is known to six loops \cite{Kompaniets:2016hct} we can calculate the first six coefficients $g_i$, $i=0,\dots,5$. Upon insertion of the six loop beta function coefficients into $g_i$, $i=0,\dots,5$ we find complete agreement with the reported results in the Mathematica file accompanying \cite{Kompaniets:2016hct}. This is a check of our formal manipulations.

\section{The $\epsilon$-Expansion as a Geometric Expansion: Associahedra}\label{associahedron}

It was expected that the $\epsilon$-expansion of the Wilson-Fisher coupling order by order is some combinatorial function of the beta function coefficients. However it has only very recently become clear to the mathematicians \cite{Loday,Loday2,Aguiar} that Lagrange inversion also has a geometric interpretation. In fact instead of viewing the arrangement of the beta function coefficients at each order in $\epsilon$ as a combinatorial exercise we can see it as being controlled by a specific polytope known as an associahedron. We note that recently the associahedron also has found its way into other branches of theoretical physics including scattering amplitudes \cite{Mizera:2017cqs,Arkani-Hamed:2017mur}.

\begin{figure}
\centering
\begin{subfigure}{.5\textwidth}
  \centering
  \includegraphics[width=.5\linewidth]{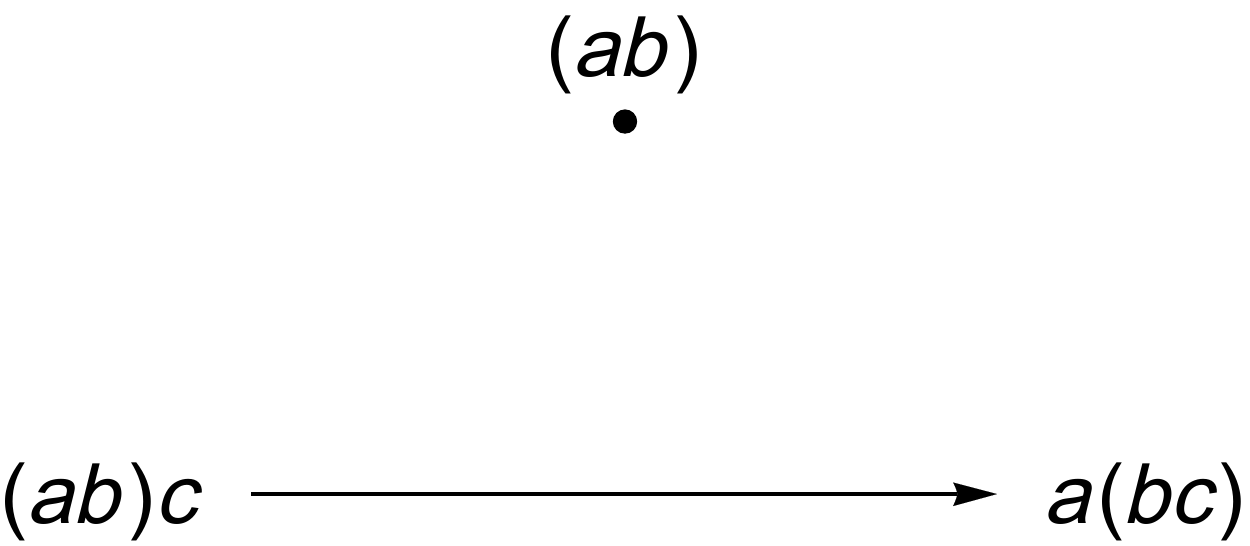}
  \caption{$K_2$ (point) and $K_3$ (line)}
  \label{fig:sub1}
\end{subfigure}%
\begin{subfigure}{.5\textwidth}
  \centering
  \includegraphics[width=.5\linewidth]{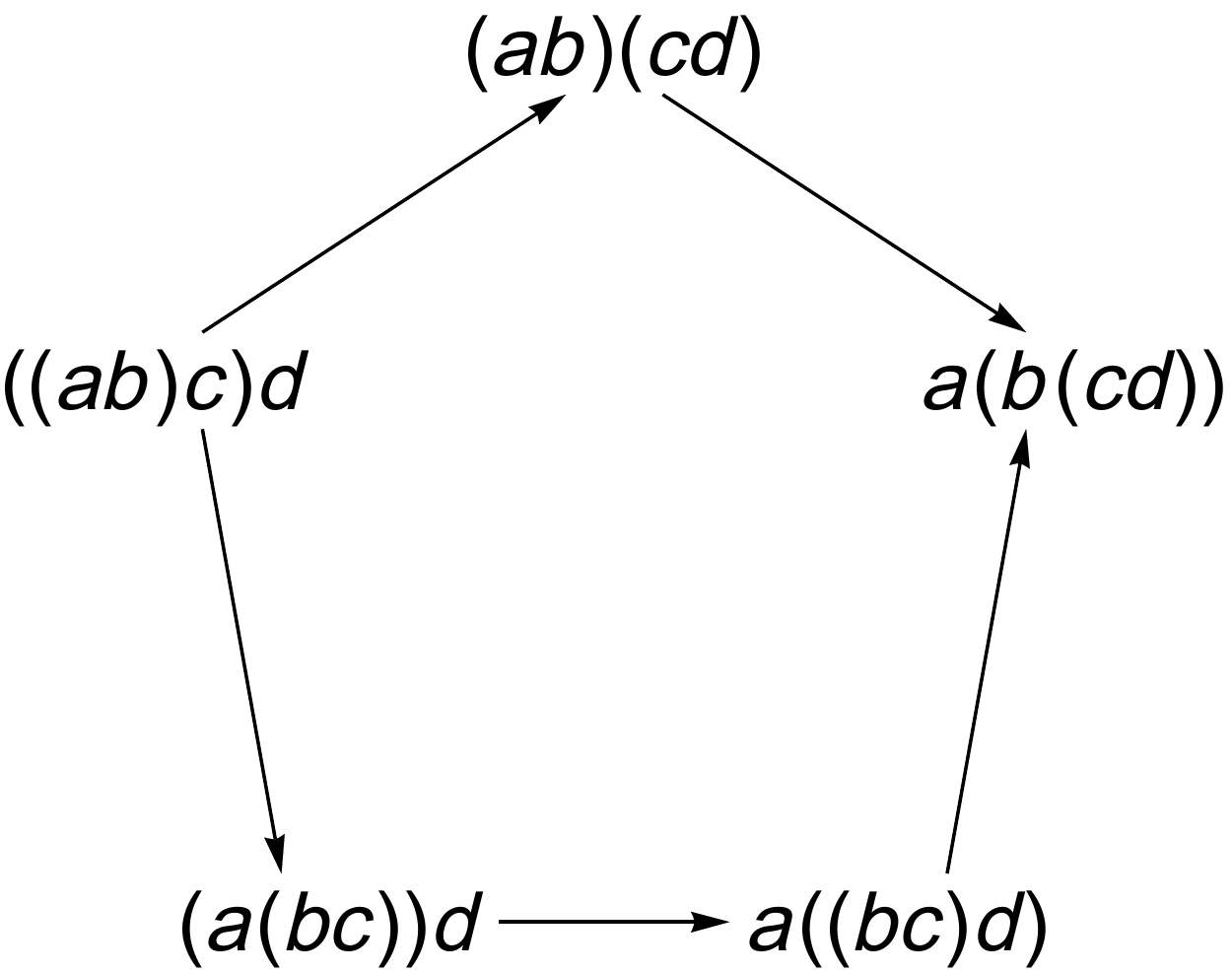}
  \caption{$K_4$ (pentagon)}
  \label{fig:sub2}
\end{subfigure}
\\
\begin{subfigure}{\textwidth}
  \centering
  \includegraphics[width=.4\linewidth]{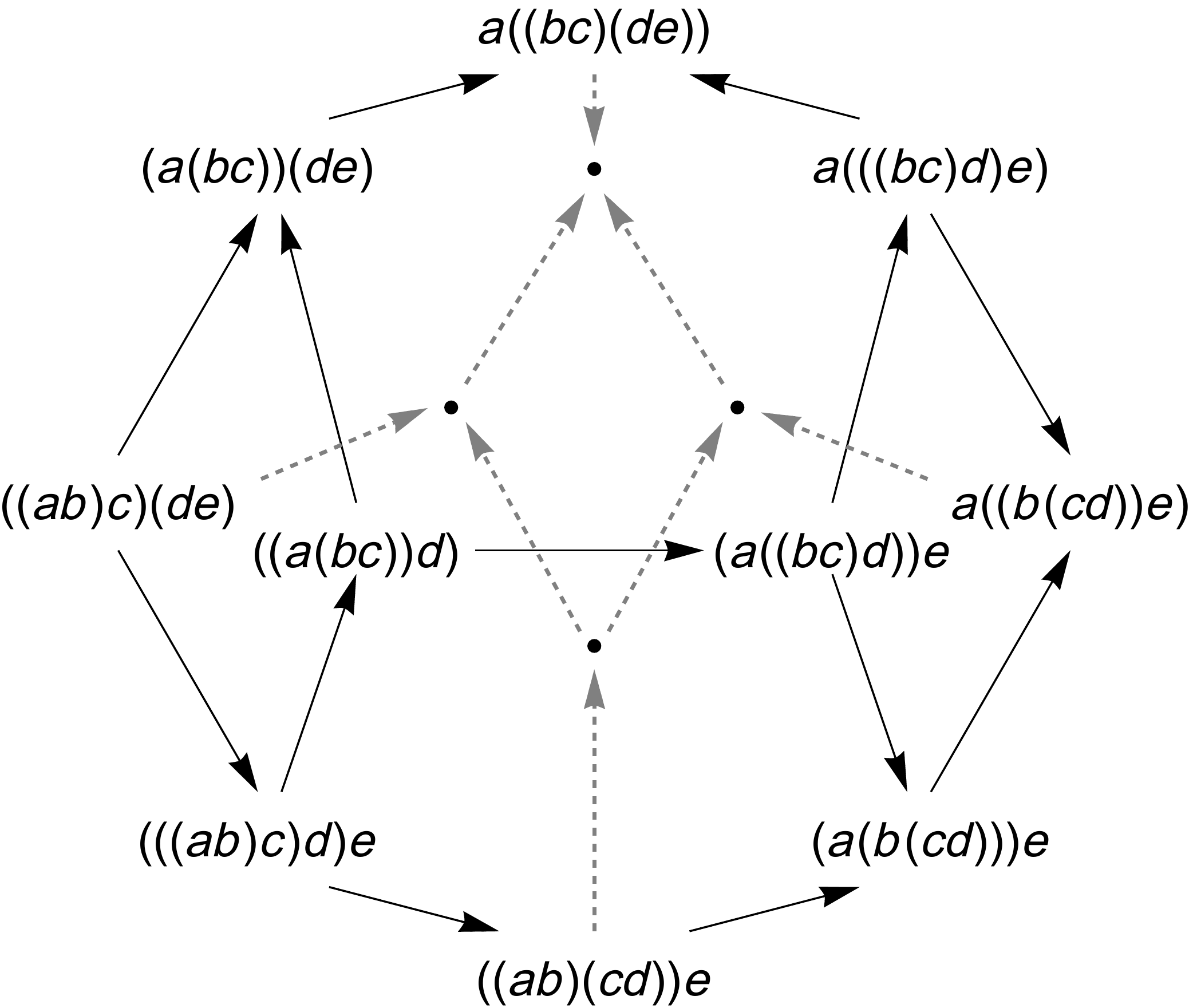}\qquad
  \includegraphics[width=.4\linewidth]{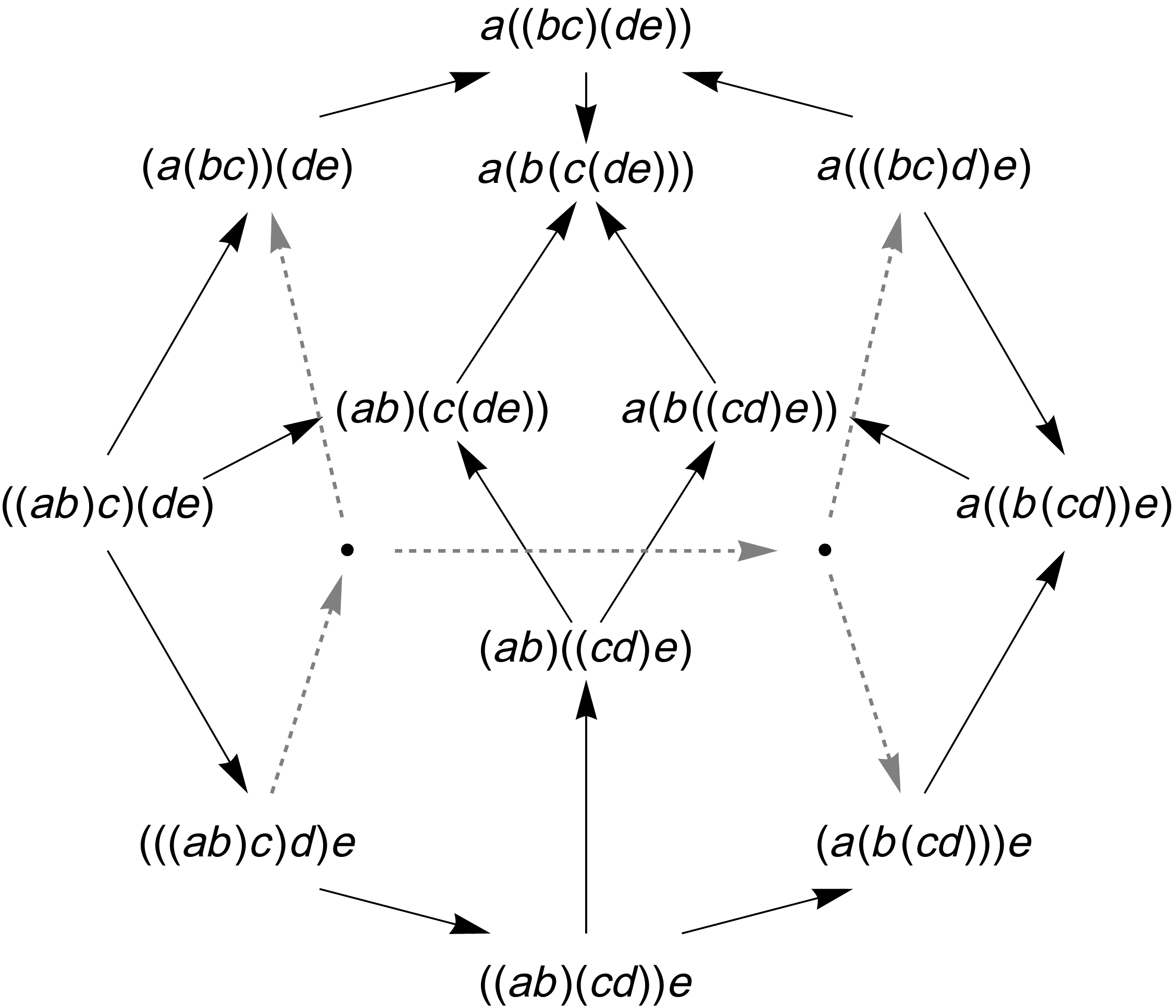}
  \caption{Front and back of $K_5$}
  \label{fig:sub3}
\end{subfigure}
\caption{The associahedra $K_2$, $K_3$, $K_4$ and $K_5$.}
\label{fig:K2K3K4K5}
\end{figure}

There are many ways to realize an associahedron \cite{Tamari,Stasheff1,Stasheff2,Loday,Loday2} (see \cite{Ziegler} for an introduction). We will define the associahedron $K_i$ as a convex polytope of dimension $i-2$. If we have a string of $i$ elements then each vertex corresponds to inserting parentheses in this string and each edge corresponds to using the associativity rule for replacing the parentheses a single time. We now construct the first few associahedra with $i=1,\ldots,5$. In Fig. \ref{fig:sub1}--\ref{fig:sub3} we show the zero, one, two and three dimensional associahedra $K_2$, $K_3$, $K_4$ and $K_5$. 

\begin{itemize}
\item If $i=1$ the associahedron $K_1$ is defined to be the empty set.
\item If $i=2$ there is only one way to put a set of parentheses in the string of $i=2$ elements $(ab)$. So the zero dimensional associahedron $K_2$ is a point.
\item If $i=3$ there are two ways to put the parentheses in the string of $i=3$ elements $(ab)c$ and $a(bc)$. So there are two vertices and the one dimensional associahedron $K_3$ is a line. 
\item If $i=4$ there are five different ways to put the parentheses in the string of $i=4$ elements $((ab)c)d$, $(ab)(cd)$, $a(b(cd))$, $a((bc)d)$ and $(a(bc))d$. So there are five vertices and the two dimensional associahedron $K_4$ is the pentagon. 
\item If $i=5$ there are fourteen ways to put the parentheses in the string of $i=5$ elements $((ab)(cd))e$, $(ab)((cd)e)$, $((ab)c)(de)$, $(ab)(c(de))$, $(a(bc))(de)$, $a((bc)(de))$, $((a(bc))d)e$, $(a((bc)d))e$, $a(((bc)d)e)$, $a((b(cd))e)$, $a(b((cd)e))$, $(a(b(cd)))e$, $(((ab)c)d)e$, $a(b(c(de)))$. So there are fourteen vertices and the three dimensional associahedron $K_5$ is composed of six pentagons and three squares. 
\end{itemize}

A few facts about the associahedron are relevant for us. Details can be found in  \cite{Tamari,Stasheff1,Stasheff2,Loday,Loday2,Ziegler}. The associahedron $K_i$ of dimension $i-2$ have faces of dimension $j$ with $j\leq i-2$. There is of course only one face of dimension $j=i-2$ which is the associahedron itself. The number of $j$-dimensional faces with $j\leq i-3$ of the associahedron $K_i$ is \cite{simion}
\begin{eqnarray}
T(i,j) = \frac{1}{i-j-1} {{i-2}\choose {i-j-2}} {{2i-j-2}\choose {i-j-2}}
\end{eqnarray}
In the number triangle in Fig. \ref{numbertriangle} we show the number of faces of different dimensions for the first few associahedra. Specifically the number of vertices $T(i,0) = \frac{1}{i} {{2i-2}\choose {i-1}}$ is known as the $(i-1)$'th Catalan number and is the leftmost column in the number triangle. Finally we need that Cartesian products of lower dimensional associahedra $F = K_{i_1} \times \cdots \times K_{i_m}$ are isomorphic to the faces of the associahedron $K_i$.  This also includes the faces which are not themselves an associahedron such as the three squares in $K_5$ which are isomorphic to $K_3\times K_3$.

\begin{figure}
 \begin{tabular}{|c|ccccc|}
 \hline
 $i$\textbackslash$j$ & 0 & 1 & 2 & 3 & 4 \\
 \hline
 1 & - & - & - & -  & - \\ 
 2 & 1 & - & - & - & - \\
 3 & 2 & 1 & - & - & -  \\
 4 & 5 & 5 & 1 & - & -  \\
 5 & 14 & 21 & 9 & 1 & -  \\
 6 &  42 & 84 & 56 & 14 & 1 \\
 \hline
\end{tabular}
\caption{Number triangle showing the number of $j$-dimensional faces $T(i,j)$ of the associahedron $K_i$. }
\label{numbertriangle}
\end{figure}

Having introduced the associahedra we now turn to how they control the $\epsilon$-expansion of the value of the coupling at the Wilson-Fisher fixed point. To each coefficient $g_{i-1}$ of the Wilson-Fisher fixed point coupling we associate the associahedron $K_i$. We will present the general result for any $i$ and then motivate it by looking at a number of examples. The calculation of the coefficient $g_{i-1}$ is controlled by the associahedron $K_i$ and is \cite{Loday,Loday2,Aguiar}
\begin{eqnarray}\label{ggeometric}
g_{i-1} =  \sum_{F\ \text{face of}\ K_i} (-1)^{i+1-\dim F}  \tilde{b}_F
\end{eqnarray} 
where  $\tilde{b}_F = \tilde{b}_{i_1-1} \cdots \tilde{b}_{i_m-1} $ for each face $F=K_{i_1}\times \cdots \times K_{i_m}$ with $i_1 +\ldots + i_m -m = i-1$  of the associahedron $K_i$. The sum is over all faces of the associahedron. In this way the facial structure of the associahedra controls the $\epsilon$-expansion of the coupling at the Wilson-Fisher fixed point. Consider now the first few examples so we can obtain a better feel for how it works.
\begin{itemize}
\item If $i=1$ the associahedron $K_1$ is the empty set and for this, by definition, we choose $g_0=1$.  
\item If $i=2$ the associahedron $K_2$ has a single face shaped like $F=K_2$ and is a point of dimension $\dim F =0$. The associated coefficient is  
\begin{eqnarray}
g_1 = (-1)^{i+1-\dim F} \tilde{b}_{2-1} =-\tilde{b}_1 
\end{eqnarray}
\item If $i=3$ the associahedron $K_3$ has two faces shaped like $F_1 = K_2 \times K_2$ which are points and each of dimension $\dim F_1 =0$, and a single face shaped like $F_2 = K_3$ which is a line and is of dimension $\dim F_2 = 1$. The associated coefficient is
\begin{eqnarray}
g_2 = 2(-1)^{i+1-\dim F_1} \tilde{b}_{2-1}\tilde{b}_{2-1} + (-1)^{i+1-\dim F_2} \tilde{b}_{3-1} = 2 \tilde{b}_1^2 - \tilde{b}_2
\end{eqnarray}
\item If $i=4$ the associahedron $K_4$ has five faces shaped like $F_1 = K_2\times K_2 \times K_2$ which are points and each of dimension $\dim F_1 = 0$, five faces shaped like $F_2 =  K_2 \times K_3$ which are lines and each of dimension $\dim F_2 = 1$ and one face shaped like $F_3 = K_4$ and of dimension $\dim F_3 = 2$. The associated coefficient is 
\begin{eqnarray}
g_3 &=& 5 (-1)^{i+1-\dim F_1} \tilde{b}_{2-1} \tilde{b}_{2-1} \tilde{b}_{2-1} + 5 (-1)^{i+1- \dim F_2} \tilde{b}_{2-1} \tilde{b}_{3-1} + (-1)^{i+1-\dim F_3} \tilde{b}_{4-1} \nonumber \\ 
&=& -5 \tilde{b}_1^3 + 5 \tilde{b}_1 \tilde{b}_2 - \tilde{b}_3 
\end{eqnarray}
\item If $i=5$ the associahedron $K_5$ has fourteen faces shaped like $F_1 = K_2 \times K_2 \times K_2 \times K_2$ which are points and of dimension $\dim F_1 = 0$, twenty one faces shaped like $F_2 =  K_2\times K_2\times K_3$ which are lines and of dimension $\dim F_2 =1$, three faces shaped like $F_3 = K_3 \times K_3$ which are squares and of dimension $\dim F_3 = 2$, six faces shaped like $F_4 =  K_2 \times K_4$ which are pentagons and of dimension $\dim F_4 = 2$ and one face shaped like $F_5 = K_5$ which is of dimension $\dim F_5 = 3$. The associated coefficient is found by the rule
\begin{eqnarray}
g_4 &=& 14 (-1)^{i+1-\dim F_1} \tilde{b}_{2-1}\tilde{b}_{2-1}\tilde{b}_{2-1}\tilde{b}_{2-1} + 21 (-1)^{i+1-\dim F_2} \tilde{b}_{2-1}\tilde{b}_{2-1}\tilde{b}_{3-1} \nonumber \\
&& + 3 (-1)^{i+1-\dim F_3} \tilde{b}_{3-1}\tilde{b}_{3-1} + 6 (-1)^{i+1-\dim F_4} \tilde{b}_{2-1}\tilde{b}_{4-1} + (-1)^{i+1-\dim F_5} \tilde{b}_{5-1} \nonumber \\
&=& 14 \tilde{b}_1^4 - 21 \tilde{b}_1^2 \tilde{b}_2 + 3 \tilde{b}_2^2 + 6 \tilde{b}_1 \tilde{b}_3 - \tilde{b}_4 
\end{eqnarray}
\end{itemize} 
In all cases is there complete agreement with our combinatorial Eq.'s \ref{g0combinatoric}--\ref{g4combinatoric}. Note that for each $i$ the sum of the indices of the beta function coefficients in $g_{i-1}$ add to $i-1$. This is a result of the condition $i_1 +\ldots + i_m -m = i-1$.

What we have arrived at is a simple and stunningly beautiful closed form expression for the coupling at the Wilson-fisher fixed point to all orders in $\epsilon$. It is dictated by the geometry of the associahedra and at each order in $\epsilon$ is uniquely related to its facial structure
\begin{eqnarray}
g_* (\epsilon) = \sum_{i=1}^{\infty} g_{i-1} \left( \frac{\epsilon}{b_0}\right)^i  \ , \qquad g_{i-1} =  \sum_{F\ \text{face of}\ K_i} \frac{1}{b_0^m} (-1)^{i+1-\dim F}  b_{i_1-1}\cdots b_{i_m-1}
\end{eqnarray}
for each of its faces $F=K_{i_1}\times \cdots \times K_{i_m}$.

\section{Anomalous Dimensions and Critical Exponents}\label{anodims}

Although the Wilson-Fisher fixed point coupling $g_*$ is scheme dependent, anomalous dimensions and critical exponents at the fixed point are not. They are scheme independent physical quantities charaterizing the system. 
  
We write the anomalous dimension of some operator in the theory as $\gamma(g)$. For instance it could be the anomalous dimension of the field $\phi$, the anomalous dimension of the mass or the anomalous dimension of some composite operator $(\phi^i\phi^i)^r$ for some integer $r$. In general the anomalous dimension is written in terms of the formal power series
\begin{eqnarray}
\gamma(g) &=& \sum_{i=1}^{\infty} \gamma_i g^i = \gamma_1 g + \gamma_2 g^2 + \gamma_3 g^3 + \ldots 
\end{eqnarray}
At the Wilson-Fisher fixed point we find to all orders in $\epsilon$
\begin{eqnarray}
\gamma(\epsilon) &=& \sum_{j=1}^{\infty} \gamma_j \left( \sum_{k=1}^{\infty} g_{k-1} \left(\frac{\epsilon}{b_0} \right)^k \right)^j = \gamma_1 \frac{\epsilon}{b_0} + \left( g_1\gamma_1 +\gamma_2 \right) \left( \frac{\epsilon}{b_0}\right)^2 + \left( g_2 \gamma_1 +2g_1 \gamma_2 +\gamma_3 \right) \left(\frac{\epsilon}{b_0} \right)^3 + \ldots \nonumber \\
&=& \sum_{i=1}^{\infty} k_i \left( \frac{\epsilon}{b_0} \right)^i \ , \qquad k_i = \frac{1}{i!} \sum_{j=1}^{i} j! \gamma_j B_{i,j}\left(1! g_0, \ldots,(i-j+1)! g_{i-j} \right) \label{eq:gamma}
\end{eqnarray}
where we have found the $i$'th coefficient $k_i$ by the same method that led us to Eq. \ref{coefficient} and the $g_{i-1}$'s are given either by the combinatoric expression Eq. \ref{gcombinatoric} or by the geometric expression Eq. \ref{ggeometric}. This constitutes an exact closed form expression for the anomalous dimension of any operator at the Wilson-Fisher fixed point. 

Note that at any order $i$ the coefficient $k_i$ only depends on the $i$ loop beta function coefficients (via $g_{i-1}$), and the coefficients of the $i$ loop anomalous dimension. It does not receive corrections from higher orders and is exact to this order. Again for the scalar $O(N)$ symmetric model using the six loop beta function, six loop field anomalous dimension $\gamma_{\phi}$ and six loop mass anomalous dimension $\gamma_{m^2}$ computed in \cite{Kompaniets:2016hct} we can compute the \emph{correction to scaling} $\omega(\epsilon) = \beta'(g_*(\epsilon),\epsilon)$ and the two \emph{critical exponents} $\eta(\epsilon) = 2 \gamma_{\phi}(g_*(\epsilon))$ and $\nu(\epsilon) = \left[ 2+ \gamma_{m^2}(g_*(\epsilon)) \right]^{-1}$ at the Wilson-Fisher fixed point to $O(\epsilon^6)$. We find complete agreement with the results reported in the Mathematica file accompanying \cite{Kompaniets:2016hct}. The same holds true for the additional critical exponents $\alpha$, $\beta$, $\gamma$ and $\delta$ related to $\eta$ and $\nu$ through the scaling relations $\gamma = \nu (2-\eta)$, $(4-\epsilon) \nu = 2-\alpha$, $\beta\delta = \beta+\gamma$ and $\alpha+2\beta+\gamma=2$. This is a final confirmation of our formal results.

\section{Scalar $O(1)$ Symmetric Model to $O(\epsilon^7)$}\label{O(1)}

We now use our general results to explicitly provide the value of the Wilson-Fisher fixed point coupling and critical exponents for the $O(1)$ model to $O(\epsilon^7)$. To our knowledge these results have still not appeared in the literature. The beta function and anomalous dimensions have been calculated to seven loops in \cite{Schnetz:2016fhy} so we can convert theses computations into a prediction of the Wilson-Fisher fixed point coupling and all the critical exponents to $O(\epsilon^7)$. Our results in this section require only little effort to arrive at and has only been made possible due to the extraordinary computations to six loops in \cite{Kompaniets:2016hct} and seven loops in \cite{Schnetz:2016fhy}. Using Eq. \ref{eq:g} and Eq. \ref{eq:gamma} they are
\begin{eqnarray}
g_*(\epsilon) &=& \frac{1}{3} \epsilon + \frac{17}{81} \epsilon^2 + \left( \frac{709}{17496} - \frac{4}{27} \zeta(3) \right) \epsilon^3   +  \left( \frac{10909}{944784} - \frac{106}{729} \zeta(3) - \frac{2}{27} \zeta(4) + \frac{40}{81} \zeta(5) \right) \epsilon^4 \nonumber \\
&& + \left( - \frac{321451}{408146688}  -  \frac{11221}{104976}\zeta(3) +   \frac{11}{81}  \zeta(3)^2 - \frac{443}{5832}\zeta(4) + \frac{373}{729} \zeta(5) + \frac{25}{54}\zeta(6) - \frac{49}{27}\zeta(7)  \right) \epsilon^5 \nonumber \\ 
&& + \left( \frac{32174329}{9183300480} - \frac{18707}{7085880} \zeta(3) + \frac{22429}{32805} \zeta(3)^2 + \frac{256}{729} \zeta(3)^3 + \frac{5776}{6075} \zeta(3,5) - \frac{19243}{314928} \zeta(4)  \right.  \nonumber \\
&&  + \frac{38}{243} \zeta(3)\zeta(4) + \frac{448}{729} \zeta(3)\zeta(5) + \frac{63481}{590490} \zeta(5)  + \frac{1117}{2187} \zeta(6) - \frac{7946}{3645} \zeta(7) - \frac{88181}{18225} \zeta(8) \nonumber \\
&&\left. + \frac{46112}{6561} \zeta(9) \right) \epsilon^6 + \left( \frac{1661059517}{991796451840} + \frac{45106}{286446699} \pi^{10} - \frac{7383787}{95659380} \zeta(3)- \frac{221281}{1180980}\zeta(3)^2  \right. \nonumber \\
&& +  \frac{19696}{19683}\zeta(3)^3  + \frac{20425591}{8266860000}\pi^8 + \frac{58}{54675} \pi^8 \zeta(3) - \frac{161678}{164025} \zeta(3,5) - \frac{112}{27} \zeta(3) \zeta(3,5)  \nonumber \\
&& - \frac{1599413}{1417176} \zeta(4) - \frac{1156}{6561} \zeta(3)\zeta(4) + \frac{16}{243} \zeta(4)^2 \frac{129631}{33067440}\pi^6 + \frac{1010}{413343} \pi^6 \zeta(3) - \frac{6227}{229635} \pi^6 \zeta(5)  \nonumber \\
&& + \frac{10590889}{85030560} \zeta(5) - \frac{163879}{19683} \zeta(3)\zeta(5) - \frac{1735}{243} \zeta(3)^2\zeta(5) - \frac{640}{729} \zeta(4)\zeta(5) + \frac{5030}{567} \zeta(5)^2  \nonumber \\
&& + \frac{12454}{1215} \zeta(5,3,3) - \frac{423301}{118098} \zeta(6) - \frac{400}{243} \zeta(3)\zeta(6) + \frac{569957}{393660} \zeta(7) + \frac{49}{1458} \zeta(3)\zeta(7) + \frac{316009}{25194240} \pi^4  \nonumber \\
&&+ \frac{4453}{393660}\pi^4 \zeta(3) + \frac{16}{2187} \pi^4 \zeta(3)^2 - \frac{613}{32805} \pi^4 \zeta(5) + \frac{6227}{18225} \pi^4 \zeta(7) - \frac{940}{5103} \zeta(7,3) \nonumber \\
&&\left.  - \frac{11992616}{492075} \zeta(8) + \frac{1347170}{177147} \zeta(9) + \frac{6227}{81} \pi^2 \zeta(9)  - \frac{8}{2187} P_{7,11} - \frac{651319}{810}\zeta(11)  \right) \epsilon^7
\end{eqnarray}
\begin{eqnarray}
\omega(\epsilon) &=& \epsilon - \frac{17}{27} \epsilon^2 + \left( \frac{1603}{2916} + \frac{8}{9} \zeta(3) \right) \epsilon^3 + \left( - \frac{178417}{314928} - \frac{158}{243} \zeta(3) + \frac{2}{3} \zeta(4) - \frac{40}{9} \zeta(5)  \right) \epsilon^4 \nonumber \\
&&+  \left( \frac{20734249}{34012224} + \frac{12349}{8748} \zeta(3) - \frac{4}{9} \zeta(3)^2 - \frac{79}{162} \zeta(4) + \frac{2324}{729} \zeta(5) - \frac{50}{9} \zeta(6) + \frac{196}{9} \zeta(7)  \right) \epsilon^5 \nonumber \\
&&+    \left( -\frac{853678429}{1224440064} - \frac{6886777}{2834352} \zeta(3) - \frac{16904}{2187} \zeta(3)^2 - \frac{1280}{243} \zeta(3)^3 - \frac{5776}{405} \zeta(3,5) + \frac{12349}{11664} \zeta(4)  \right. \nonumber \\
&& - \frac{2}{3} \zeta(3)\zeta(4) - \frac{95713}{39336} \zeta(5) -  \frac{4960}{243} \zeta(3) \zeta(5) +  \frac{5405}{1458} \zeta(6) - \frac{961}{81} \zeta(7) + \frac{88181}{1215} \zeta(8) \nonumber \\
&& \left. - \frac{230560}{2187} \zeta(9)  \right) \epsilon^6 + \left( \frac{99202757785}{132239526912}  - \frac{316009}{1399680} \pi^4  - \frac{129631}{1837080} \pi^6 - \frac{20425591}{459270000} \pi^8   \right. \nonumber \\
&&  -\frac{90212}{31827411} \pi^{10} \zeta(11) + \frac{480656027}{102036672} \zeta(3) - \frac{4453}{21870} \pi^4 \zeta(3) - \frac{2020}{45927} \pi^6 \zeta(3) - \frac{116}{6075} \pi^8 \zeta(3)      \nonumber \\
&& + \frac{1737593}{78732} \zeta(3)^2 - \frac{32}{243} \pi^4 \zeta(3)^2 - \frac{64312}{6561} \zeta(3)^2+ \frac{508228}{10935} \zeta(3,5) + \frac{224}{3} \zeta(3)  \zeta(3,5) + \frac{34951705}{1889568} \zeta(4)   \nonumber \\
&& + \frac{4907}{729} \zeta(3) \zeta(4) - \frac{16}{27} \zeta(4)^2 + \frac{198223}{314928} \zeta(5)+ \frac{1226}{3645} \pi^4 \zeta(5) + \frac{12454}{25515} \pi^6 \zeta(5) + \frac{385046}{2187} \zeta(3) \zeta(5)   \nonumber \\
&& + \frac{3470}{27} \zeta(3)^2 \zeta(5) + \frac{640}{81} \zeta(4) \zeta(5) - \frac{226820}{1701} \zeta(5)^2 - \frac{24908}{135} \zeta(5,3,3) + \frac{10053541}{157464} \zeta(6)  \nonumber \\
&& + \frac{1300}{81} \zeta(3) \zeta(6)  - \frac{755965}{26244} \zeta(7) - \frac{12454}{2025} \pi^4 \zeta(7) + \frac{1421}{27} \zeta(3) \zeta(7) + \frac{1880}{567}\zeta(7,3)   \nonumber \\
&&\left. + \frac{47970464}{164025} \zeta(8) + \frac{4459444}{59049} \zeta(9) - \frac{12454}{9} \pi^2 \zeta(9) +\frac{16}{243} P_{7,11} \right) \epsilon^7
\end{eqnarray}
\begin{eqnarray}
\eta(\epsilon) &=& \frac{1}{54} \epsilon^2 + \frac{109}{5832} \epsilon^3+\left(  \frac{7217}{629856}  -\frac{4}{243} \zeta(3) \right) \epsilon^4 + \left( \frac{321511}{68024448} - \frac{329}{17496} \zeta(3) - \frac{1}{84} \zeta(4) + \frac{40}{729} \zeta(5)  \right) \epsilon^5 \nonumber \\
&& \left( \frac{46425175}{264479053824} - \frac{4247}{25194240} \pi^4 - \frac{71}{1180980} \pi^6 - \frac{2063}{229635000} \pi^8- \frac{1978411}{204073344} \zeta(3)  \right.   \nonumber \\
&&- \frac{1}{21870}\pi^4 \zeta(3)  + \frac{10027}{157464}  \zeta(3)^2 + \frac{256}{6561} \zeta(3)^3 + \frac{244}{2187} \zeta(3,5)  + \frac{11969}{3779136} \zeta(4) +\frac{22}{729} \zeta(3) \zeta(4) \nonumber \\
&&\left. + \frac{59917}{1889568} \zeta(5) + \frac{50}{729} \zeta(3) \zeta(5)   + \frac{42397}{314928} \zeta(6) - \frac{3707}{17496} \zeta(7) - \frac{88181}{164025} \zeta(8) + \frac{46112}{59049} \zeta(9)  \right) \epsilon^7 \nonumber \\
\end{eqnarray}
\begin{eqnarray}
\nu(\epsilon) &=&\frac{1}{2}+ \frac{1}{12} \epsilon + \frac{7}{162} \epsilon^2 + \left( \frac{1783}{69984} - \frac{1}{27}\zeta(3)  \right) \epsilon^3 + \left( \frac{92969}{7558272}  -\frac{191}{5832} \zeta(3) - \frac{1}{36} \zeta(4) + \frac{10}{81} \zeta(5) \right) \epsilon^4 \nonumber \\
&&+  \left( \frac{4349263}{816293376} - \frac{6323}{209952} \zeta(3) + \frac{2}{81} \zeta(3)^2 - \frac{191}{7776} \zeta(4)   + \frac{74}{729} \zeta(5) + \frac{25}{162} \zeta(6) - \frac{49}{108}  \zeta(7)  \right) \epsilon^5   \nonumber \\
&&\left(  \frac{65712521}{29386561536}  -\frac{58565}{2519424}\zeta(3)  + \frac{2807}{26244} \zeta(3)^2 + \frac{64}{729} \zeta(3)^3  +  \frac{61}{243} \zeta(3,5) - \frac{6323}{279936} \zeta(4) \right. \nonumber \\
&& + \frac{1}{27} \zeta(3) \zeta(4) + \frac{132893}{1889568} \zeta(5) + \frac{184}{729}  \zeta(3) \zeta(5) + \frac{1615}{11664} \zeta(6) - \frac{4255}{11664} \zeta(7) - \frac{16337}{11664} \zeta(8) \nonumber \\
&& \left. + \frac{11528}{6561} \zeta(9) \right) \epsilon^6 + \left( \frac{3466530079}{3173748645888} +\frac{312061}{100776960} \pi^4 + \frac{463493}{396809280} \pi^6 + \frac{7085207}{11022480000} \pi^8 \right.  \nonumber \\
&&+ \frac{22553}{477411165} \pi^{10}  - \frac{651319}{3240} \zeta(11) -\frac{53182423}{2448880128} \zeta(3) + \frac{35}{11664} \pi^4 \zeta(3) + \frac{79}{91854} \pi^6 \zeta(3) \nonumber \\
&& + \frac{29}{109350} \pi^8 \zeta(3) + \frac{244339}{3779136}  \zeta(3)^2 + \frac{8}{3645} \pi^4 \zeta(3)^2 + \frac{3991}{19683} \zeta(3)^3 - \frac{13633}{131220} \zeta(3,5) \nonumber \\
&&  - \frac{28}{27} \zeta(3) \zeta(3,5) - \frac{47}{5103} \zeta(3,7) - \frac{248687}{839808} \zeta(4) - \frac{959}{8748} \zeta(3) \zeta(4) + \frac{2}{81} \zeta(4)^2 + \frac{3664579}{68024448} \zeta(5) \nonumber \\
&& - \frac{1393}{262440} \pi^4 \zeta(5) - \frac{6227}{918540} \pi^6 \zeta(5)  - \frac{23827}{19683} \zeta(3) \zeta(5) -  \frac{1735}{972} \zeta(3)^2 \zeta(5) - \frac{200}{729} \zeta(4) \zeta(5) \nonumber \\
&& + \frac{26008}{15309} \zeta(5)^2 + \frac{6227}{2430} \zeta(5,3,3) - \frac{3800527}{3779136} \zeta(6) - \frac{725}{1458} \zeta(3) \zeta(6) - \frac{1951}{157464} \zeta(7)  + \frac{6227}{72900} \pi^4 \zeta(7) \nonumber \\
&& \left. - \frac{35}{972} \zeta(3) \zeta(7) - \frac{235}{5103} \zeta(7,3) - \frac{52036931}{7873200} \zeta(8) + \frac{2654189}{2834352} \zeta(9) + \frac{6227}{324}  - \frac{2}{2187} P_{7,11}  \right) \epsilon^7
\end{eqnarray}
where
\begin{eqnarray}
\zeta(k) = \sum_{n=1} \frac{1}{n^k} \ , \qquad \zeta(k_i,\dots,k_1) \sum_{n_i>\ldots>n_1 \geq 1} \frac{1}{n_i^{k_i} \cdots n_1^{k_1}}
\end{eqnarray}
and $P_{7,11}$ was calculated in \cite{Panzer:2015ida}. Numerically  it is $P_{7,11} = 200.357566....$. We also evaluate the fixed point coupling and critical exponents numerically. They are
\begin{eqnarray}
g_*(\epsilon) &=& 0.333333\epsilon +0.209877 \epsilon^2 - 0.137559 \epsilon^3 + 0.268653 \epsilon^4 - 0.843685 \epsilon^5 +3.15437 \epsilon^6 \nonumber \\ 
&& -13.4831 \epsilon^7  \\
\omega(\epsilon) &=& \epsilon  - 0.62963 \epsilon^2 + 1.61822 \epsilon^3 - 5.23514 \epsilon^4 + 20.7498 \epsilon^5 - 93.1113 \epsilon^6 + 458.742 \epsilon^7 \\
\eta(\epsilon) &=& 0.0185185\epsilon^2 + 0.01869 \epsilon^3 - 0.00832877 \epsilon^4 + 0.0256565 \epsilon^5 - 0.0812726 \epsilon^6  + 0.314749 \epsilon^7 \\
\nu(\epsilon) &=& 0.5 + 0.083333 \epsilon + 0.0432099 \epsilon^2  - 0.0190434 \epsilon^3 + 0.0708838 \epsilon^4 - 0.217 018 \epsilon^5 + 0.829313 \epsilon^6 \nonumber \\
&& - 3.57525 \epsilon^7
\end{eqnarray}

\section{Discussion}\label{conclusions}

In this work we have used Lagrange inversion to derive the exact form of the Wilson-Fisher fixed point coupling and critical exponents in the $\epsilon$-expansion in terms of the coefficients of the appropriate renormalization group functions. We have also argued that the $\epsilon$-expansion of the Wilson-Fisher fixed point coupling can be viewed in a geometric sense and is controlled by the associahedra.

Although we explicitly discussed the scalar $O(n)$ symmetric model in $d=4-\epsilon$ dimensions in the introduction our analysis is certainly not restricted to this set of theories. In fact in the above derivations it could be any theory with a single coupling and for which $b_0 >0$ so that the Wilson-Fisher fixed point appears for $d<4$, ($\epsilon>0$). One could also imagine extending our analysis to multiple couplings. Also for theories for which $b_0<0$ one could perform similar studies as above for $d>4$, ($\epsilon<0$). In this way we envision many new projects worth investigating. We should mention that our analysis is independent of whether the various different renormalization group functions have a finite radius of convergence or are asymptotic. This is irrelevant to the Lagrange inversion procedure as given here order by order. 

One of our original hopes was that in deriving a closed form expression order by order for the Wilson-Fisher fixed point coupling we would gain further insight into the asymptotic nature of the expansion. Whether we view the expansion as combinatoric in terms of Bell polynomials or as geometric in terms of the facial structure of associahedra this hope however has unfortunately not been fulfilled. Nevertheless we believe that our results are useful since one at least has an understanding of how the various beta function coefficients enter at a given order for example.

Finally we used our general results to compute the Wilson-Fisher fixed point coupling and the associated critical exponents to $O(\epsilon^7)$ for the $O(1)$ model. We also gave the numerical values.

{\flushleft\textbf{Acknowledgments:}} We thank G. Dunne, E. M\o lgaard and R. Shrock for important discussions. TAR is partially supported by the Danish National Research Foundation under the grant DNRF:90.

\bibliographystyle{apsrev4-1}

\end{document}